\documentclass[10pt, conference, compsocconf]{IEEEtran}
\usepackage{amsmath}
\usepackage{amssymb}
\usepackage{graphicx}
\usepackage{cite}
\usepackage{color}

\usepackage{subfig}
\usepackage[ruled]{algorithm2e}
\SetKwInput{KwInit}{Initialize}
\SetKwInput{KwInput}{Input}
\SetKwInput{KwOutput}{Output}
\SetKwFor{parfor}{parfor}{:}{end parfor}

\usepackage{url}

\usepackage{MnSymbol}
\usepackage{listings}
\lstdefinelanguage{scala}{
  morekeywords={abstract,case,catch,class,def,%
    do,else,extends,false,final,finally,%
    for,if,implicit,import,match,mixin,%
    new,null,object,override,package,%
    private,protected,requires,return,sealed,%
    super,this,throw,trait,true,try,%
    type,val,var,while,with,yield},
  otherkeywords={=>,<-,<\%,<:,>:,\#,@},
  sensitive=true,
  morecomment=[l]{//},
  morecomment=[n]{/*}{*/},
  morestring=[b]",
  morestring=[b]',
  morestring=[b]"""
}
\newcommand{\specialcell}[2][c]{%
  \begin{tabular}[#1]{@{}c@{}}#2\end{tabular}}

\ifCLASSINFOpdf
\else
\fi
\begin{document}
\title{Geotagging One Hundred Million Twitter Accounts with Total Variation Minimization}


\author{\IEEEauthorblockN{Ryan Compton, David Jurgens, David Allen}
\IEEEauthorblockA{Information and System Sciences Laboratory\\
HRL Laboratories\\
3011 Malibu Canyon Rd, Malibu, CA 90265.\\
rfcompton@hrl.com}
}


%Telephone: (800) 555--1212, Fax: (888) 555--1212}


\maketitle

\begin{abstract}
	Geographically annotated social media is extremely valuable for modern information retrieval. However, when researchers can only access publicly-visible data, one quickly finds that social media users rarely publish location information. In this work, we provide a method which can geolocate the overwhelming majority of active Twitter users, independent of their location sharing preferences, using only publicly-visible Twitter data.
	
	%The goal of this work is to demonstrate a technique for large-scale location inference in Twitter data.
	%We motivate our use of total variation --- which leads to a network where connected users are often placed into the same city --- 
	
	Our method infers an unknown user's location by examining their friend's locations. We frame the geotagging problem as an optimization over a social network with a total variation-based objective and provide a scalable and distributed algorithm for its solution. Furthermore, we show how a robust estimate of the geographic dispersion of each user's ego network can be used as a per-user accuracy measure which is effective at removing outlying errors.

	Leave-many-out evaluation shows that our method is able to infer location for $101,846,236$ Twitter users at a median error of $6.38$ km, allowing us to geotag over $80\%$ of public tweets.
\end{abstract}


% \fi
\IEEEpeerreviewmaketitle

\section{Introduction}
\label{sec:introduction}

The ability to geospatially index a large volume of Twitter data is valuable for several emerging research directions. Indeed, geographic social media analytics have proven useful for understanding regional flu trends \cite{paul2011you}, linguistic patterns \cite{deliaBabel}, election forecasting \cite{tumasjan2010predicting}, social unrest \cite{rcompton2013seattle}, and disaster response \cite{mandel2012demographic}. These approaches, however, depend on the physical locations of Twitter users which are only sparsely available in public data. %\footnote{Less than $1\%$ of tweets are annotated with GPS and less than 30\% of users report unambiguous locations in their profiles.}

Interestingly, recent work from the computational social sciences community has established that online social ties are often formed over short geographic distances \cite{takhteyev2012geography} \cite{mok2010does} \cite{goldenberg2009distance}. Because of this, it is possible to approximate the location of a Twitter user by examining publicly-known locations of their online friends \cite{DavidJ} \cite{Yamaguchi:2013:LUL:2512938.2512941}. Using social network analysis to solve geolocation problems relies only on public Twitter metadata, which, somewhat counterintuitively, provides several advantages over content-based approaches. Network-based geotagging sidesteps difficulties with foreign natural language processing, ignores noisy Twitter text, and makes it possible to demonstrate results at previously unreachable scales. The largest and most accurate geolocation results currently published have utilized social network analysis \cite{DavidJ} \cite{backstrom2010find}.

In this work, we solve network-based geotagging problems from an optimization viewpoint.  Unlike existing methods which have independently developed node-wise heuristics, we show that a globally-defined and highly-studied convex optimization can be solved for location inference. To be precise, we infer user location by solving
\begin{equation}
\label{eq:tvopt}
\min_{\mathbf{f}} | \nabla \mathbf{f} | \; \text{subject to} \; f_i = l_i \; \text{for} \; i \in L
\end{equation}
where $\mathbf{f} = (f_1,\ldots,f_n )$ encodes a location estimate for each user, $L$ denotes the set of users who opt to make their locations, $l_i$, public, the total variation on the Twitter social network is defined by
\begin{equation}
\label{eq:tvdef}
| \nabla \mathbf{f} | = \sum_{ij} w_{ij} d(f_i,f_j)
\end{equation}
where $d(\cdot, \cdot)$ measures geodesic distance via Vincenty's formulae, and the edge weights, $w_{ij}$, are equal to the minimal number of reciprocated @mentions between users $i$ and $j$.

In other words, \textbf{we seek a network such that the weighted sum over all geographic distances between connected users is as small as possible}. This sum, defined in (\ref{eq:tvdef}) and known in the literature as total variation, has demonstrated superior performance as an optimization heuristic for several information inference problems across a wide variety of fields.

The minimization in (\ref{eq:tvopt}) leads to a network where connected communities of users are placed into the same city. Motivation for this heuristic relies on communicative locality in Twitter which has been demonstrated in previous research \cite{takhteyev2012geography} \cite{DavidJ} \cite{Yamaguchi:2013:LUL:2512938.2512941} and is further justified by our experiments in section \ref{sec:grnd}. Evidence of communicative locality in other online social networks can be found in \cite{mok2010does} \cite{goldenberg2009distance} and \cite{backstrom2010find}. 

The social media geotagging problem is relatively new; total variation minimization, however, is not. Originally introduced to the computational sciences as a heuristic for image denoising \cite{Rudin1992}, literature on total variation minimization is now truly vast --- thousands of papers spanning several decades. While most of this research has been confined to imaging processing, recent work has shown that total variation is valuable to several newer applications, such as crime modelling \cite{Smith:2010:IDE:1840693.1928520}, graphics \cite{osher2003geometric}, and ``community detection'' \cite{hu2013method}. Very recently, it has been shown that use of total variation for general transductive learning tasks outperforms state-of-the-art methods on standard benchmark datasets \cite{bresson2013multiclass} \cite{bresson2012convergence}. The present work is the first to demonstrate a theory link between the social media geotagging problem and total variation minimization.

There are real and immediate practical applications for our results. From a business perspective, the importance of Twitter geotagging has recently warranted several commercial offerings\footnote{\url{http://blog.gnip.com/twitter-geo-data-enrichment}}\footnote{\url{http://tweepsmap.com}}. At least one product\footnote{\url{http://support.gnip.com/enrichments/profile_geo.html}} is based on alignments between self-reported profile locations and the GeoNames\footnote{\url{http://geonames.org}} database. Our experiments in section \ref{sec:coverage} demonstrate that, even if all nonempty profile locations could be mapped to unambiguous locations, the coverage of profile-based geotagging could reach at most $63\%$ of tweet volume. When nonsensical or ambiguous profile locations are ignored, our experiments show that the coverage of profile-based geotagging drops below $20\%$.

A salient feature of our approach, however, is that we are unable to estimate within-city motion for individual Twitter users. Our goal is high-volume static location inference with city-level accuracy. This may be tolerable for several applications as research has indicated that users typically remain within a small radius; cf. section \ref{sec:motion} and 
\cite{chen2013interest}. Since it is still unclear if better-than-city-level accuracy is possible at scale, we restrict our training sets to users who tweet primarily within the same radius, using their median location as ``home''%. (cf. \ref{sec:grnd}).% We justify this decision with data on Twitter user mobility in \ref{sec:motion}.

Using social networks to infer location makes sense only if a user's friends are primarily located within the same geographic region. We show that the geographic dispersion of each user's ego network\footnote{the ``ego network'' of a user consists of the user together with the users they are directly connected to},
\begin{equation}
\label{eq:mediantv}
\overset{\sim}{\nabla f_i} = \text{median}_j d(f_i,f_j)
\end{equation}
agrees well with geotag error and therefore provides us with a per-user confidence measure. Overall error can thus be controlled by limiting $\max \overset{\sim}{\nabla f_i}$. Our experiments show that coverage remains high even when restrictions on $\max \overset{\sim}{\nabla f_i}$ are tight.

Solutions to the geotagging problem are most interesting when demonstrated at scale. We exhibit our method on a network of $110,893,747$ Twitter users with $1,034,362,407$ connections between them and infer locations for $101,846,236$ users. This is made possible by the optimization algorithm we employ to solve (\ref{eq:tvopt}), parallel coordinate descent \cite{richtarik2012parallel}, which we have implemented in a distributed manner using the Apache Spark cluster computing framework \cite{zaharia2012resilient}.  

To be clear, the contributions of this paper are as follows:
\begin{itemize}
\item{We show that the social network geotagging problem can be solved via a globally-defined convex optimization}
\item{We develop a novel per-user accuracy estimate}
\item{We demonstrate our results at scale; producing the largest database of Twitter user locations currently known in the literature}
\end{itemize}

\section{Background and Related Work}
\label{sec:background}

Location inference in social media has caught the attention of several researchers from both academia and industry. Two major classes of solutions to the geotagging problem are now prevalent: language-based, and network-based.

\subsection{Language-based Geotagging} 
Work by \cite{cheng2010you} and \cite{eisenstein2010latent} provided methods based on identifying and searching for location-specific terms in Twitter text. More recent work by \cite{mahmud2012tweet} (extended in \cite{mahmudhome}) builds on these approaches with ensemble classifiers that account for additional features such as time zone and volume of tweets per hour. To the best of our knowledge, \cite{mahmudhome} showcases the current state-of-the-art for content-based geotagging: $9,551$ test users at $68\%$ city-accuracy. Our goal is similar to that of \cite{mahmudhome} in that both works focus on static location inference with city-resolution accuracy, however, we demonstrate several orders of magnitude greater coverage as well as higher accuracy. Specifically, our leave-many-out validation tests indicate that we can correctly predict the city a user resides in with vs $89.7\%$ city-accurate vs. the $68\%$ reported by \cite{mahmudhome}.

Geotagging via natural language processing requires that users from different geographic regions tweet in different dialects, and that these differences are great enough to make accurate location inference possible. Evidence against this possibility appears in \cite{DavidJ}, where the author examined a large Twitter dataset and found minimal agreement between language models and proximity. Additionally, language-based geotagging methods often rely on sophisticated language-specific natural language processing and are thus difficult to extend worldwide.

\subsection{Network-based Geotagging}
Large-scale language-agnostic geotagging is possible by inferring a user's location with the known locations of their friends. Influential work in this field was conducted at Facebook Inc. in \cite{backstrom2010find}. Here, the authors infer a user's home address with a maximum likelihood estimate that best fits an empirically observed power law. Surprisingly, the results of \cite{backstrom2010find} indicate that social network based methods are \textit{more} accurate than IP-based geolocation. Since the IP addresses of Twitter users are never public, and our research involves public data alone, we can not report any results about how the present work compares against IP-based Twitter geolocation.

Geotagging work on Flickr and Twitter by Sadilek \textit{et al.} \cite{sadilek2012finding} uses social ties to infer location and then studies the converse problem: using location to infer social ties. They conclude that location alone is insufficient for this task. Very recent work by \cite{Yamaguchi:2013:LUL:2512938.2512941} uses network structure as well as language processing of user profiles to identify ``landmark'' users in the United States for whom location inference is optimal while \cite{Ryoo:2014:ITU:2567948.2579236} showcases a similar result on Korean Twitter users.

Most closely related to our work is that of \cite{DavidJ}, where the author developed a node-wise algorithm, ``Spatial Label Propagation'', which iteratively estimates Twitter user locations by propagating the locations of GPS-known users across a Twitter social network. While not discussed in \cite{DavidJ}, it turns out that Spatial Label Propagation is in fact a parallel coordinate descent method applied to total variation minimization. Later in this work we show how our technique can reduce to Spatial Label Propagation by removing constraints on (\ref{eq:mediantv}). While Spatial Label Propagation was demonstrated at scale, our study reaches higher coverage and accuracy: $101.8$M users at a median error of $6.33$km vs. $45.8$M users at a median error of $10$km. 

Research indicating that Twitter contact is independent of proximity can be found in \cite{leetaru2013mapping}, where the author examines GPS-known retweet pairings and finds an average distance of $749$ miles between users. Averages, however, are sensitive to outliers which are often present in social data. In this work, we will make use of robust statistics to estimate center and spread for sets of locations.

\section{Method}

\subsection{Data and Network Construction}
\label{sec:graph}
An appropriate social network is a fundamental part of our algorithm. Twitter users often ``@mention'' each other by appending an ``@'' to the mentioned user's name. We build a social network, $G=(V,E)$, with users as vertices and @mentions between users as edges.

Reciprocated @mentions indicate social ties. We define edge weights, $w_{ij}$, of $G$, the ``bidirectional @mention network'', using the minimum number of reciprocated @mentions between users $i$ and $j$. The key advantage to constructing a social graph from @mentions (as opposed from ``followers" or ``favourites") is that it enables us to build a large social graph from a large collection of tweets without being burdened by Twitter API rate limiting.

We use a $10\%$ sample of public tweets collected between April 2012 and April 2014. This amounts to $76.9$TB of json data (uncompressed) and $25,312,399,718$ @mentions. From the complete set of @mentions we built a weighted and directed network of $8,593,341,111$ edges by condensing multiple mentions into weighted edges. Filtering down to only reciprocated @mentions leaves us with a network of $\mathbf{ 1,034,362,407}$ edges\footnote{i.e. a $12\%$ chance of @mention reciprocation} and $\mathbf{110,893,747}$ users. This bidirected network is the focus of our experiments.

\subsection{Ground Truth User Locations}
\label{sec:grnd}

We define a function, $f$, which assigns to each user an estimate of their physical location. Users may opt to make their location publicly available though cellphone-enabled GPS or self-reported profile information. For this small set of users, computation of $f$ is relatively straightforward.

To assign a unique location to a user from the set of their GPS-tagged tweets, $\mathcal{G}$, we compute the $l1$-multivariate median \cite{vardi2000multivariate} of the locations they have tweeted from:
\begin{equation}
\underset{x}{\operatorname{\text{argmin}}} \; \sum_{y \in \mathcal{G}} d(x,y)
\end{equation}

Our data contains $13,899,315$ users who have tweeted with GPS at least three times. Several of these users are highly mobile and can not reasonably be assigned a single, static, location. We filter out users whose median absolute deviation (cf. (\ref{eq:mediantv})) of GPS-annotated tweet locations is over $30$km. This leaves with a set of $12,435,622$ users whose location might be known via GPS. However, tweet timestamps reveal that $86,243$ of these users have at some point exceeded the flight airspeed record of $3529.6$ km/h. Manual examination of these accounts finds several bots retweeting worldwide GPS-annotated tweets as well as human users who suffer GPS malfunctions (e.g. a tweet near $(0.0,0.0)$ shortly after a valid location). We remove from our training set any user who has travelled in excess of $1,000$ km/h. The total number of GPS-known users is $12,297,785$.

Following this, we extract self-reported home locations by searching through a list of $51,483$ unambiguous location names for exact matches in user profiles. Users who list several locations in their profile are not geotagged by this step, though it is possible to account for such users using a method found in \cite{Yamaguchi:2013:LUL:2512938.2512941}. When self-reporting users also reveal their location though GPS, we opt to use their GPS-known location. We remove self-reports which are over $90$ days old. This provides us with home locations for an additional $15,360,494$ users.

\begin{figure}
\centering
\includegraphics[width=\columnwidth]{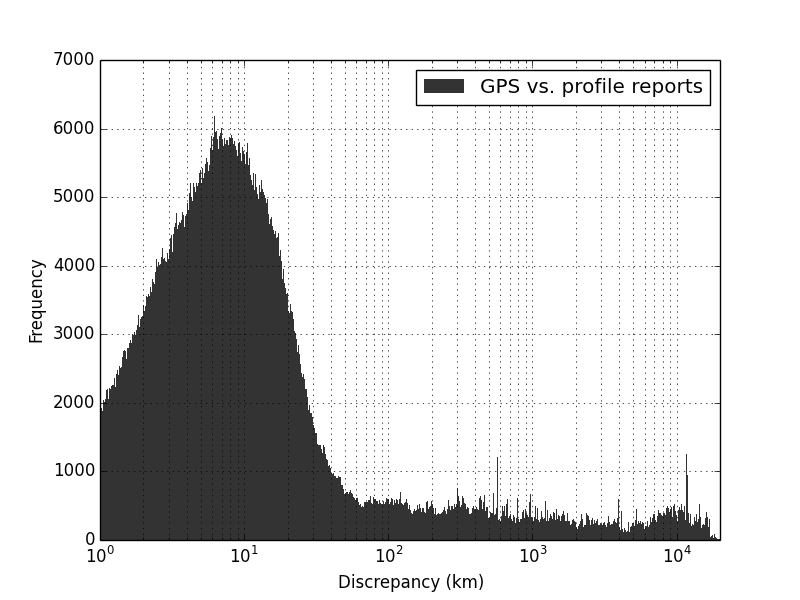}
\caption{Histogram of discrepancies between GPS and self-reported profile locations for $1,883,331$ users. The median and mean discrepancies are $7.10$ and $318.50$ km, respectively.
}
\label{fig:self_report_discrep_per_tweet}
\end{figure}

The list of $51,483$ location names has been optimized for accuracy on Twitter. Starting with an initial list of $67,711$ location names obtained from the GeoNames project, we examined $12,471,920$ GPS-tagged tweets\footnote{collected between 2013-01-01 and 2013-04-07} from users who self-reported profile locations and removed location names from the list when the median discrepancy between GPS and the reported location was greater than $100$km. We have plotted the resulting discrepancies between GPS and self-reports in fig. \ref{fig:self_report_discrep_per_tweet}.

The total number of users with ground truth locations via GPS or self-reports is $\mathbf{24,545,425}$. Denote this set of users by $L$, and the remaining users in the network by $U$. The vertex set of our social network is thus partitioned as
\begin{equation}
V = L + U
\end{equation}
and our goal is to assign a value of $f$ to nodes in $U$.

\subsection{Global Optimization Algorithm}
\label{sec:opt}

%figure that gene stanley likes
%the * makes it 2-column
\begin{figure*}
\centering
\subfloat[gpsmin][\label{mention_ecdf} Minimum distance to a friend ]{
\includegraphics[width=.31\textwidth]{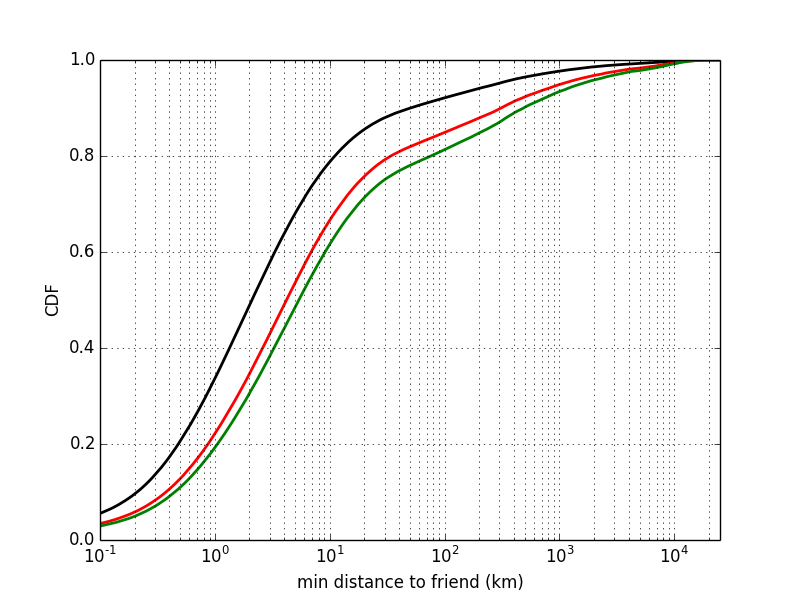}
}
\subfloat[gpsmedian][\label{closest_mention_ecdf} Median distance to a friend]{
\includegraphics[width=.31\textwidth]{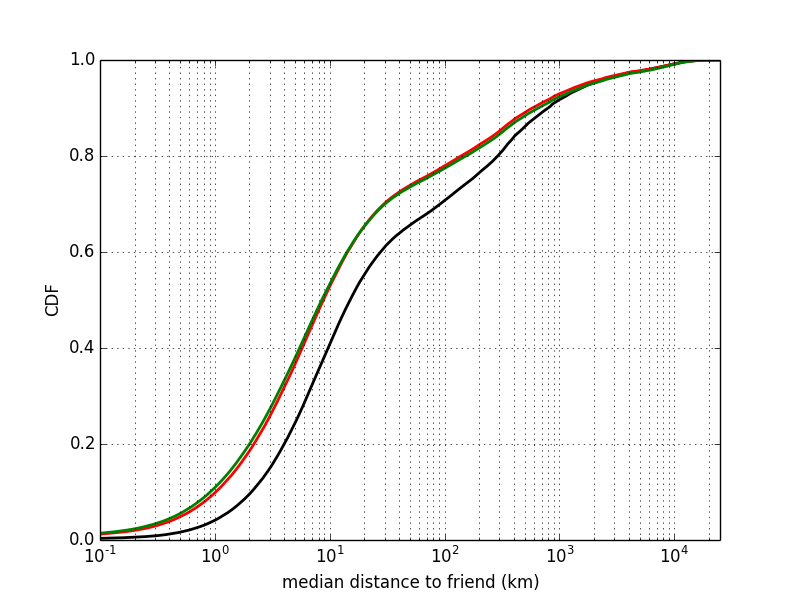}
}
\subfloat[gpsmax][\label{mention_ecdf} Maximum distance to a friend]{
\includegraphics[width=.31\textwidth]{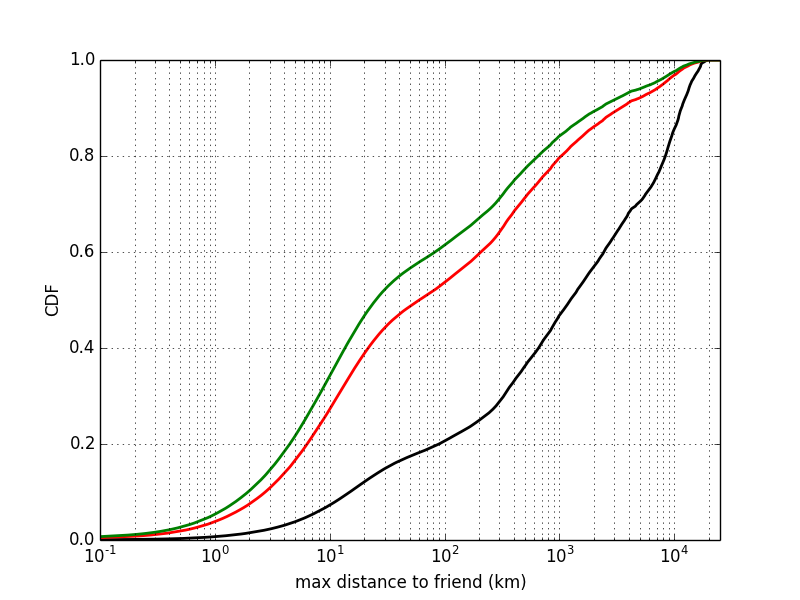}
}
\\
\caption{\label{gps_vs_edges}
Study of contact patterns between $953,557$ users who reveal their location via GPS and are present in each of the bidirectional @mention network (red), bidirectional @mention network after filtering edges for triadic closures (green), and the unfiltered unidirectional @mention network (black).
}
\end{figure*}
Our algorithm assigns a location to a user based on the locations of their friends. To check that online social ties are well-aligned with geographic distance, we restrict our attention to GPS-known users and study contact patterns between them in fig. \ref{gps_vs_edges}.

Users with GPS-known locations make up only a tiny portion of our @mention networks. Despite the relatively small amount of data, we can still see in fig. \ref{gps_vs_edges} that online social ties typically form between users who live near each other and that a majority of GPS-known users have at least one GPS-known friend within 10km.

The optimization (\ref{eq:tvopt}) models proximity of connected users. Unfortunately, the total variation functional is nondifferentiable and finding a global minimum is thus a formidable challenge. We will employ ``parallel coordinate descent'' \cite{richtarik2012parallel} to solve (\ref{eq:tvopt}). Most variants of coordinate descent cycle through the domain sequentially, updating each variable and communicating back the result before the next variable can update. The scale of our data necessitates a parallel approach, prohibiting us from making all the communication steps required by a traditional coordinate descent method.

At each iteration, our algorithm simultaneously updates each user's location with the $l1$-multivariate median of their friend's locations. Only after all updates are complete do we communicate our results over the network.
  
At iteration $k$, denote the user estimates by $\mathbf{f}^k$ and the variation on the $i$th node by
\begin{equation}
\label{l1med2}
|\nabla_i(\mathbf{f}^k,f)| = \sum_j w_{ij} d(f,f^k_j)
\end{equation}
Parallel coordinate descent can now be stated concisely in alg. \ref{alg:pcd2}.

% \begin{algorithm}
% \KwInit{$f_{i} = l_{i}$ for $i\in L$}
% \For{$k = 1 \ldots N$}{
% \parfor{$i$}{
% \eIf{$i \in L$}{
% $f^{k+1}_i = l_i$
% }{
% $f^{k+1}_i = \underset{f}{\operatorname{\text{argmin}}} |\nabla_i(\mathbf{f}^k,f)|$
% }
% }
% $\mathbf{f}^{k} = \mathbf{f}^{k+1}$
% }
% \caption{Parallel coordinate descent for constrained TV minimization}
% \label{alg:pcd}
% \end{algorithm} 

\subsection{Individual Error Estimation}

The vast majority of Twitter users @mention with geographically close users. However, there do exist several users who have amassed friends dispersed around the globe. For these users, our approach should not be used to infer location.

We use a robust estimate of the dispersion of each user's friend locations to infer accuracy of our geocoding algorithm. Our estimate for the error on user $i$ is the median absolute deviation of the inferred locations of user $i$'s friends, computed via (\ref{eq:mediantv}). With a dispersion restriction as an additional parameter, $\gamma$, our optimization becomes
\begin{equation}
\label{eq:tvopt2}
\min_{\mathbf{f}} | \nabla \mathbf{f} | \; \text{subject to} \; f_i = l_i \; \text{for} \; i \in L \; \text{and} \; \max_i \overset{\sim}{\nabla f_i} < \gamma 
\end{equation}

\begin{algorithm}
\KwInit{$f_{i} = l_{i}$ for $i\in L$ and parameter $\gamma$}
\For{$k = 1 \ldots N$}{
\parfor{$i$}{
\eIf{$i \in L$}{
$f^{k+1}_i = l_i$
}{
	\eIf{$\overset{\sim}{\nabla f_i} \leq \gamma$}
		{$f^{k+1}_i = \underset{f}{\operatorname{\text{argmin}}} |\nabla_i(\mathbf{f}^k,f)|$}
		{no update on $f_i$}
}
}
$\mathbf{f}^{k} = \mathbf{f}^{k+1}$
}
\caption{Parallel coordinate descent for dispersion-constrained TV minimization.}
\label{alg:pcd2}
\end{algorithm}

Restricting the maximum allowed value of $\overset{\sim}{\nabla f_i}$ during each update ensures that only reliable locations are propagated through subsequent iterations. In alg. \ref{alg:pcd2} we refuse to update locations for users whose friends are dispersed beyond a given threshold.

The argument that minimizes (\ref{l1med2}) is a weighted $l1$-multivariate median of the locations of the neighbors of node $i$. By placing this computation inside the \textbf{parfor} of alg. \ref{alg:pcd2} and removing any restriction on $\gamma$, we are able to reproduce the ``Spatial Label Propagation'' algorithm of \cite{DavidJ} as a coordinate descent method designed to minimize total variation.

While several researchers have worked with parallel coordinate descent, existing convergence results are difficult to apply. This is in part due to the fact that most convergence studies assume Euclidean space while our algorithms involve metrics on the surface of a sphere. We also face the problem of ensuring that the ground truth users fully retain their initial location for all iterations, which prohibits us from working with an $l2$-penalized unconstrained problem as is often studied in theoretical papers.

\subsection{Implementation Remarks}

\begin{lstlisting}[caption=Distributed implementation of alg. \ref{alg:pcd2} using the Spark framework \cite{zaharia2012resilient},
  label=lst:pcd,
  float=t]
val edgeList = loadEdgeList()
var userLocations = loadInitialLocations()    
for (k <- 1 to N) {
  val adjListWithLocations =  edgeList.join(userLocations) .keyBy(x => x._2).groupByKey()
  val updatedLocs = adjListWithLocations.map(x => (x._1, l1Median(x._2), dispersion(x._2)))
  userLocations = updatedLocs.filter(x => x._3 < GAMMA)							
}
return userLocations   		 
\end{lstlisting}

The Apache Spark cluster computing framework \cite{zaharia2012resilient} was used to implement alg. \ref{alg:pcd2}. Spark allows one to distribute data in cluster memory by making use of resilient distributed datasets (referred to as RDDs) and operate on these datasets with arbitrary Scala code.

Our technique is sketched in Listing \ref{lst:pcd}. Our network and user locations are stored in the RDDs \texttt{edgeList}, and \texttt{userLocations}. Computations on these RDDs make use of all available cluster cpu resources. The \textbf{parfor} of alg. \ref{alg:pcd2} can be implemented with a straightforward \texttt{map} and \texttt{filter}. Communicating the updated locations across the network is accomplished with a \texttt{join} on the edge list, followed by a \texttt{groupByKey}, which sets up an adjacency list for the next \texttt{map}.

An implementation of alg. \ref{alg:pcd2} taking advantage of more advanced distributed graph processing techniques, such as the Pregel model \cite{malewicz2010pregel} or GraphX \cite{xin2013graphx}, is a direction for future work.

\subsection{Mobility Considerations}
\label{sec:motion}

\begin{figure}
\centering
\includegraphics[width=\columnwidth,height=0.75\columnwidth]{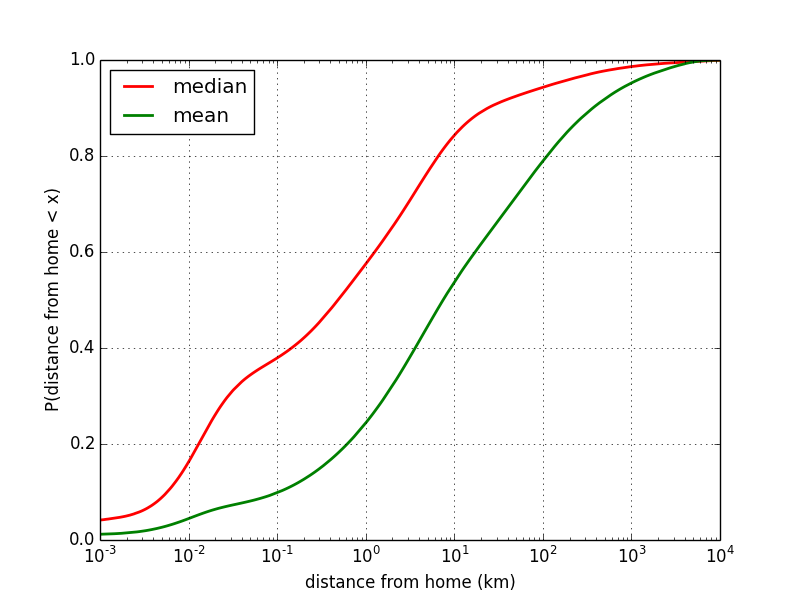}
\caption{Empirical CDF of activity radii for GPS-known users. For each of the $13,899,315$ users with $3$ or more GPS-annotated tweets, we compute their home location using the median of their tweet locations and define activity radii using the mean and median distances from home.}
\label{fig:gps-spread-cdf2}
\end{figure}

The technique outlined above is only useful for static location inference. Fast-moving users with large activity radii will be tagged incorrectly by our method. Here, we study the appropriateness of static user geotagging by restricting our attention to GPS-annotated tweets and reporting statistics on activity radii of these users.

For the $13,899,315$ users with three or more GPS-annotated tweets (cf. section \ref{sec:grnd}), we define a home location using the median of their tweet locations and plot empirical cumulative distribution functions describing the mean and median distances from home in fig. \ref{fig:gps-spread-cdf2}. The data indicates that large activity radii exist, but are atypical of Twitter users.

The presence of Twitter users with large activity radii is expected. Several recent quantitative studies of mobility patterns in public social media have confirmed that distance from home is decisively fails to follow a normal distribution. Power law, lognormal \cite{mocanu2013twitter}, gravity law \cite{hawelka2013geo}, and radiation laws have been studied. What this means for our research is that outliers are unavoidable in geographic social data and the use of robust statistics is a must.  

Large activity radii correspond to faster moving users. Timestamps on the GPS-annotated tweets reveal the median speed of a Twitter user is $.02$ km/h (average speed $233.63$ km/h). For users with an activity radius over $20$km, the median jumps an order of magnitude to $0.20$ km/h (average speed $1024.04$ km/h). Fast-moving accounts may be inhuman and not geotaggable. For example, the maximum speed attained by any Twitter user in our data was $67,587,505.24$ km/h, slightly more than $30$x the escape velocity from the surface of the Sun. 

%The median distance from home for these accounts is $1.05$ km (mean distance $1806.60$ km), a result which helps us emphasize the dispersion constraints we placed on the ground truth users in \ref{sec:grnd}.

\section{Results}
\label{sec:results}

We run our optimization on the bidirectional @mention network described in section \ref{sec:graph}. Results are reported after $5$ iterations of alg. \ref{alg:pcd2}, though high coverage can be obtained obtained sooner (cf. table \ref{tbl:num_users_geo_per_iter}). The parameter $\gamma$ was set to $100$km after experimenting with different values (cf. fig. \ref{fig:error_histogram_trunc} and fig. \ref{fig:gamma_lines}). 

\subsection{Coverage}
\label{sec:coverage}

To assess coverage, we examined $37,400,698,296$ tweets collected between April 2012 and April 2014. These tweets were generated by $359,583,211$ users.

\begin{table}
\centering
\resizebox{\columnwidth}{!}{ 
    \begin{tabular}{lllll}
    Iteration & Test users & \specialcell{Test users\\added} & \specialcell{Median \\ error (km)} & \specialcell{Median error on\\new test users} \\ \hline
    1         & 771,321    & 771,321          & 5.34              & 5.34                                \\
    2         & 926,019    & 154,698          & 6.02              & 12.31                               \\
    3         & 956,705    & 30,686           & 6.24              & 45.50                               \\
    4         & 966,515    & 9,810            & 6.32              & 150.60                              \\
    5         & 971,731    & 5,216            & 6.38              & 232.92                              \\
    \end{tabular}
}
\caption{Geolocated users and accuracy for each iteration. The first iterations produce the most accurate geotags and the highest coverage.}
\label{tbl:num_users_geo_per_iter}
\end{table}

\begin{figure}
\centering
\includegraphics[width=\columnwidth]{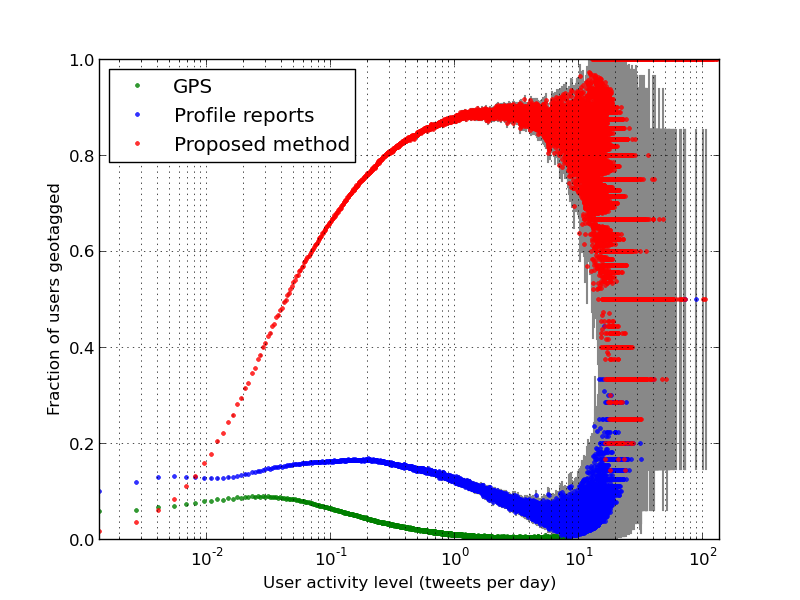}
\caption{Fraction of geolocated users as a function of activity level. The probability that a user reveals their location via GPS (green) or self-reports (blue) changes little. The probability that a user is geotagged by our method (red) is dramatically higher for users who tweet more often. Error bars (grey) display the standard error of the mean at each activity level and are largest for high activity levels (where the number of sample points is small).}
\label{fig:n_users_per_tweet_count_hrl_ratio}
\end{figure}

\begin{figure}
\centering
\includegraphics[width=\columnwidth,height=0.75\columnwidth]{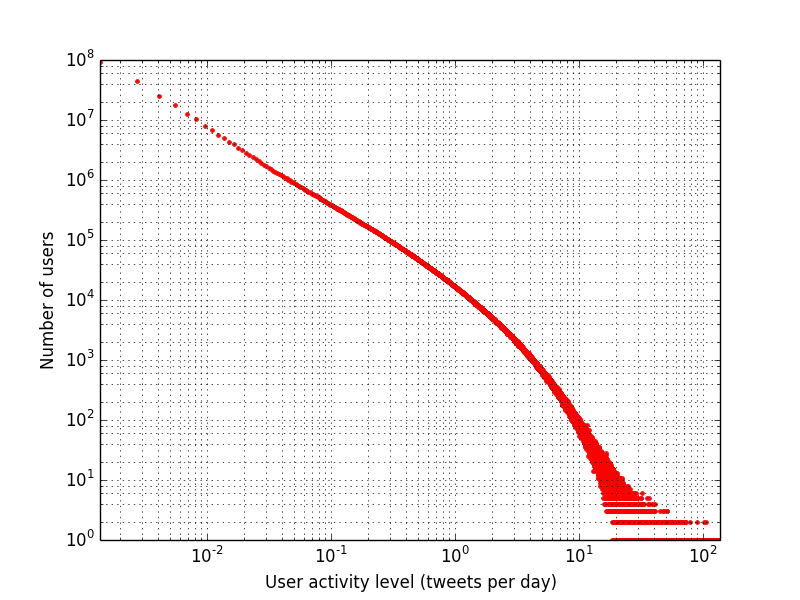}
\caption{Log-log plot describing the total number of users as a function of activity level. A large fraction of users quickly lose interest in Twitter after generating a small number of tweets. As observed in fig. \ref{fig:n_users_per_tweet_count_hrl_ratio}, these users are difficult to geocode with the proposed method.}
\label{fig:n_users_per_tweet_count}
\end{figure}

Active Twitter users are likely present in the @mention network and are therefore likely geocoded by our algorithm. In fig. \ref{fig:n_users_per_tweet_count_hrl_ratio}, we grouped users by activity level and, for each group, plotted the probability that a user is geolocated by three different methods. The probability that a user is geocoded by our method increases dramatically as a function activity level. The probability that a user is geocoded via GPS or self-reports, however, appears to be unrelated to activity level. We summarize coverage results in table \ref{table:n_users_per_tweet_count_dallen}.

The statistics in fig. \ref{fig:n_users_per_tweet_count_hrl_ratio} become noisy for extremely active users. This is due to the fact that only a small number of users tweet this often (cf fig. \ref{fig:n_users_per_tweet_count}). Given that we collect only $10\%$ of Twitter, activity levels over $10^1$ in our chart contain users who have been tweeting over $100$ times per day over a two-year period which may indicate that they are not human. 

\begin{table}
\centering
\resizebox{\columnwidth}{!}{ 
    \begin{tabular}{lll}

    ~                     & Tweet volume              & Account volume         \\ \hline
    GPS                   & 584,442,852 (1.6\%)    & 12,297,785 (3.4\%)  \\
    \specialcell{unambiguous\\profile}   & 3,854,169,186 (10.3\%)  & 45,284,996 (12.6\%) \\
    \specialcell{any nonempty\\profile} & 23,236,139,825 (62.1\%)  & 164,020,169 (45.6\%) \\
    proposed method       & 30,617,806,498 (81.9\%)  & 101,846,236 (28.3\%) \\
    \end{tabular}
}
    \caption{Summary of coverage results. The sample data consisted of $37,400,698,296$ tweets generated by $359,583,211$ users.  We are able to geotag $81.9\%$ of tweets, more than is possible even if all nonempty self-reported profile locations were unambiguous.}
\label{table:n_users_per_tweet_count_dallen}
\end{table}

\subsection{Accuracy}

Accuracy is assessed using leave-many-out cross-validation with a $10\%$ hold out set. From the $12,297,785$ users who reveal GPS locations, we randomly selected $1,229,523$ test users for exclusion from $L$. After $5$ iterations of alg. \ref{alg:pcd2} with $\gamma = 100$km, we were able to infer location for $971,731$ test users with a median error of $6.38$km and a mean error of $289.00$km.

Reverse-geocoding to cities with a population over $5,000$ shows that our method was accurate to city-resolution for $770,498$ $(89.7\%)$ of test users. We remark here that evaluations based on semantic distance (which are common in language-based geotagging) can be difficult to compare with those based on physical distance as reverse-geocoding can introduce errors of its own. For example, depending on the convention used, a minimum population size of $5,000$ would discriminate between the 48 different barrios of Buenos Aires which are physically close yet semantically distinct.

\begin{figure}
\centering
\includegraphics[width=\columnwidth]{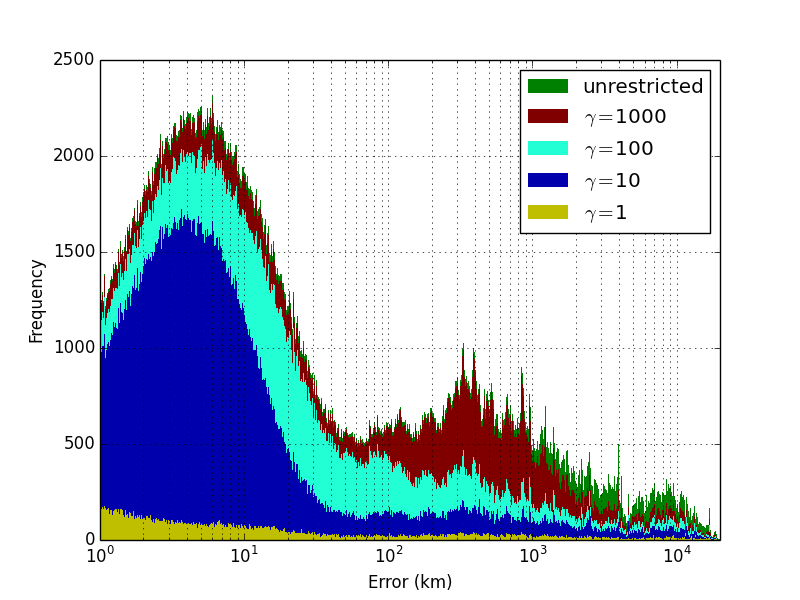}
\caption{Histogram of errors with different restrictions on the maximum allowable geographic dispersion of each user's ego network in km (i.e. $\gamma$ in (\ref{eq:tvopt2}) ). We are able to remove outlying errors by controlling $\gamma$.}
\label{fig:error_histogram_trunc}
\end{figure}

\begin{figure}
\centering
\includegraphics[width=\columnwidth]{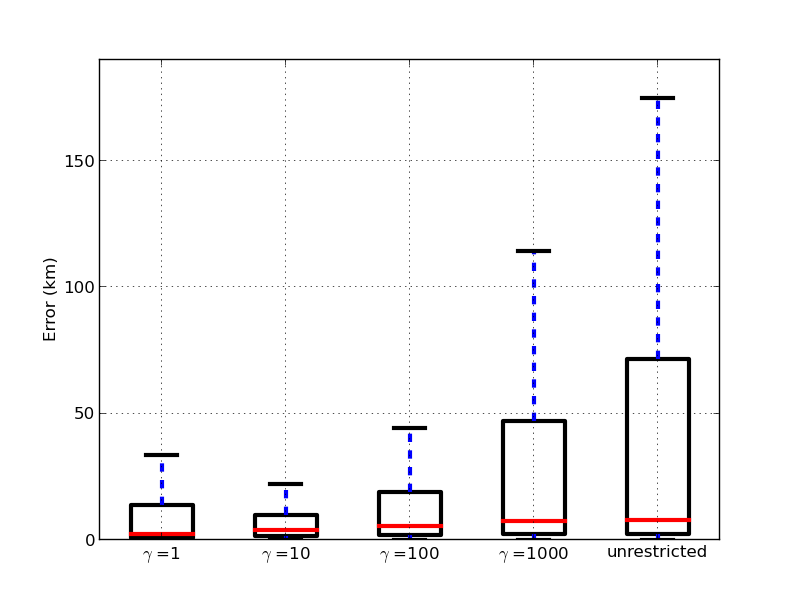}
\caption{Box plots describing the errors for different values of $\gamma$. The ends of the whiskers are drawn at $\pm 1.5$ times the interquartile range.}
\label{fig:boxplots}
\end{figure}

Inferring location for users with geographically dispersed ego networks leads to large errors. In fig. \ref{fig:error_histogram_trunc} we use data from a run with no restriction on $\max \overset{\sim}{\nabla f_i}$ (a.k.a Spatial Label Propagation) and observe bimodal error distributions when restrictions on $\max \overset{\sim}{\nabla f_i}$ are loose. When we account for ego network dispersion with $\gamma$ in (\ref{eq:tvopt2}) we are able to eliminate outlying errors. The effect of modifying $\gamma$ is further illustrated with box plots in In fig. \ref{fig:boxplots}.

 \begin{figure}
 \centering
 \subfloat[mean-rroc][\label{sfig:mean-rroc} Mean error control via $\gamma$]{
 \includegraphics[width=\columnwidth,height=0.6\columnwidth]{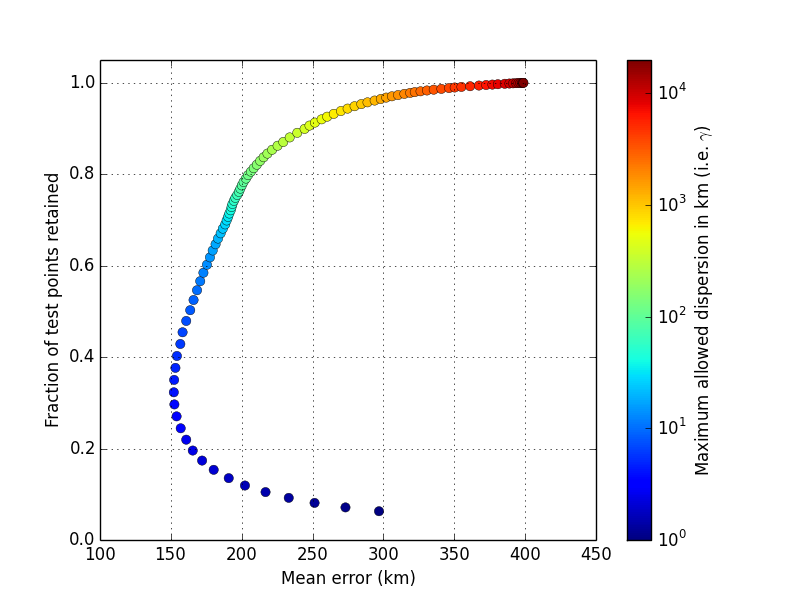}
 }\\
 \subfloat[median-rroc][\label{sfig:median-rroc} Median error control via $\gamma$]{
 \includegraphics[width=\columnwidth,height=0.6\columnwidth]{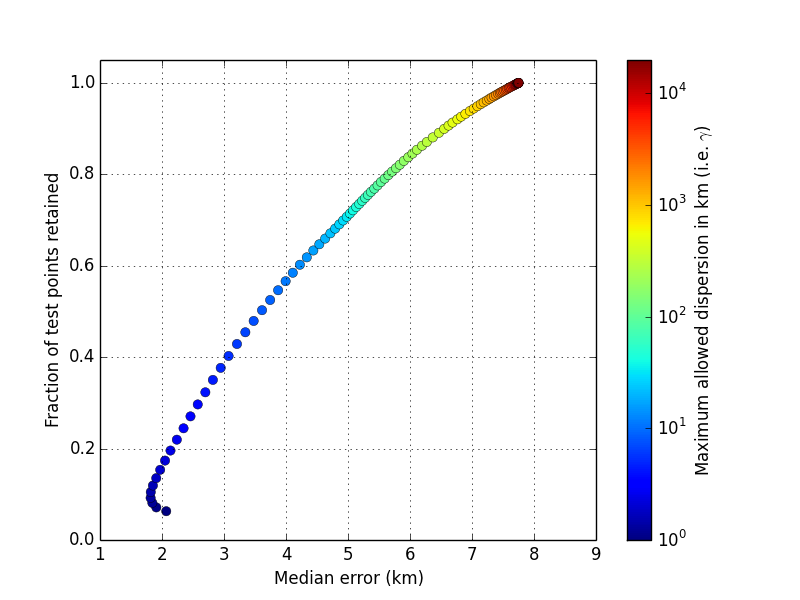}
 }\\
 \caption{\label{fig:gamma_lines}
 Coverage penalties and accuracy gains as a function of $\gamma$. In fig. \ref{sfig:mean-rroc} we observe that a restriction of $\gamma=100$km reduces mean error to $200$km while retaining $80\%$ of users in our geolocation database. In fig. \ref{sfig:median-rroc} we see that median error is less responsive to $\gamma$, indicating that restrictions on $\gamma$ primarily control outlying errors.}
\end{figure}

%\begin{figure}
%\centering
%\includegraphics[width=\columnwidth]{mean_err_scatter.png}
%\caption{Coverage penalties and accuracy gains as a function of $\gamma$. With a restriction of $\gamma=100$km we retain close to $80\%$ of users and reduce the mean error by $50\%$. }
%\label{fig:gamma_lines}
%\end{figure}

Coverage is not substantially decreased by restricting ego network dispersion. Our experiments with tuning $\gamma$ are visible in fig. \ref{fig:gamma_lines} where we observe that the fraction of geolocated users remains high even when $\gamma$ is made small.

\section{Conclusion}

We have presented a total variation-based algorithm for inferring the home locations of millions of Twitter users. By framing the social network geotagging problem as a global convex optimization, we have connected much recent work in computational social science with an immense body of existing knowledge.

Additionally, we have developed a novel technique to estimate per-user accuracy of our geotagging algorithm and used it to ensure that errors remain small. 

Our results have been demonstrated at scale. To the best of our knowledge, this is the largest and most accurate dataset of Twitter user locations known. The fact that it is constructable from publicly visible data opens the door for many future research directions.

\section*{Acknowledgment}

Supported by the Intelligence Advanced Research Projects Activity (IARPA) via Department of Interior National Business Center (DoI / NBC) contract number D12PC00285. The U.S. Government is authorized to reproduce and distribute reprints for Governmental purposes notwithstanding any copyright annotation thereon. The views and conclusions contained herein are those of the authors and should not be interpreted as necessarily representing the official policies or endorsements, either expressed or implied, of IARPA, DoI/NBE, or the U.S. Government.



%\bibliographystyle{IEEEtran}

\end{document}